\documentclass[twocolumn, superscriptaddress]{revtex4-1}
\usepackage{amssymb}
\usepackage{amsmath}
\usepackage{epsfig}
\usepackage{color}
\usepackage{graphics, graphicx}
\usepackage{bbold}
\usepackage{psfrag}
\usepackage{mathcomp}
\usepackage{subfigure}
\usepackage{verbatim}
\usepackage{color}
\usepackage[colorlinks,citecolor=blue]{hyperref}
\def\cp#1{\mathbf{#1}}

\begin{document}
\title{Enhanced fermion pairing and superfluidity by an imaginary magnetic field }
\author{Lihong Zhou}
\affiliation{Beijing National Laboratory for Condensed Matter Physics, Institute of Physics, Chinese Academy of Sciences, Beijing 100190, China}
\author{Xiaoling Cui}
\email{xlcui@iphy.ac.cn}
\affiliation{Beijing National Laboratory for Condensed Matter Physics, Institute of Physics, Chinese Academy of Sciences, Beijing 100190, China}
\affiliation{Songshan Lake Materials Laboratory , Dongguan, Guangdong 523808, China}
\date{\today}

\begin{abstract}
We show that an imaginary magnetic field(IMF), which can be generated in non-Hermitian systems with spin-dependent dissipations, can greatly enhance the s-wave pairing and superfluidity of spin-1/2 fermions, in distinct contrast to the effect of a real magnetic field.  The enhancement can be attributed to the increased coupling constant in low-energy space and the reduced spin gap in forming singlet pairs. We have demonstrated this effect in a number of different fermion systems with and without spin-orbit coupling,  using both the two-body exact solution and many-body mean-field theory. Our results suggest an alternative route towards strong fermion superfluid with high superfluid transition temperature.
\end{abstract}

\maketitle

{\bf Introduction.}

Searching for strong fermion superfluids and their underlying mechanisms has been one of the central tasks in  condensed matter and cold atomic physics. In cold atoms, a prominent example for strong superfluid is the unitary Fermi gas, where the s-wave scattering length diverges and the interaction energy solely scales with the Fermi energy\cite{FermiGas}. Apart from resonant interaction, several other factors have also been shown to induce strong pairing and superfluidity, such as low dimension\cite{2D}, 
large effective range\cite{narrow},  highly-symmetric spin-orbit coupling\cite{SOC}, etc. On the contrary, the presence of a magnetic field or spin imbalance is generally believed to reduce and even destroy the pairing superfluidity, especially when the spin gap overcomes the pairing strength.

In this work, we report another efficient tool for generating strong pairing superfluid, namely, an imaginary magnetic field (IMF).  Experimentally, the IMF can be realized in non-Hermitian atomic systems by laser-assisted spin-selective dissipations\cite{dissipation_expt}.
Consider the spin-1/2($\uparrow,\downarrow$) system, the IMF can be equivalently achieved by applying a laser field uniquely to spin-$\downarrow$ atom, which is resonantly coupled to a highly excited atomic state and causes loss. This spin-dependent loss can be described by a potential $i\Gamma \sigma_z$ up to a constant energy shift ($\sim -i\Gamma/2$), where $\Gamma$ determines the loss rate. Such potential exactly plays the role of an imaginary Zeeman energy due to an imaginary magnetic field $B=i\Gamma$.
 Here we show that the IMF can greatly enhance the s-wave pairing and superfluidity of spin-1/2 fermions,  behaving just oppositely to a real magnetic field(RMF). Consider a Rashba spin-orbit coupled (SOC) fermion system as an example, we find that even a small IMF can induce an exponential enhancement of the two-body binding energy in weak coupling regime, and the enhancement equally holds for the 
pairing superfluid of many fermions in all interaction regime.  
We further demonstrate that the IMF-enhanced pairing commonly exists in several other typical fermion systems, with different types of SOC and even without SOC. 
The enhancement can be attributed to the increased coupling constant in low-energy space and the reduced spin gap in forming singlet pairs when an IMF is present. These results, which are detectable in current cold atoms experiment, suggest an alternative route towards strong fermion superfluid with high superfluid transition temperature.

{\bf Results and Discussion.}

To demonstrate the effect of an IMF, we start with a concrete model of spin-1/2 fermions($\uparrow,\downarrow$) with Rashba SOC. The single-particle Hamiltonian in momentum (${\cp k}$) space can be written as ($\hbar=1$ throughout the paper)
\begin{equation}
h_0(\cp{k})=\epsilon_{\cp{k}}\sigma_0+\alpha(k_x\sigma_x+k_y\sigma_y)+B\sigma_z, \label{h0}
\end{equation}
here $\epsilon_{\cp k}={\cp k}^2/(2m)$; $\sigma_0$ and $\sigma_{x,y,z}$ are respectively the identity and Pauli matrices; $\alpha$ is the strength of Rashba SOC, which naturally defines a momentum scale $k_0=2m\alpha$ and an energy scale $E_0=2m\alpha^2$; $B$ can be real or imaginary, respectively denoting a RMF or an IMF. The eigen-energies of (\ref{h0}) are
\begin{equation}
\xi_{\cp k,\pm}=\epsilon_{\cp k}\pm\sqrt{\alpha^2k_{\perp}^2+B^2}, \label{E_single}
\end{equation}
where $\pm$ is the helicity index and $k_\perp=\sqrt{k_x^2+k_y^2}$. 
Note that in writing Eq.\ref{h0} with an imaginary $B$, we have utilized the effective non-Hermitian Hamiltonian reduced from the Lindblad equation, by neglecting a term that induces quantum jumps between diagonal density matrixes in different particle-number sectors. 
It has been argued that such process will not affect the physical quantities produced with a given particle number\cite{Ueda-Yu}. 

\begin{figure}[t]
\includegraphics[width=8.5cm]{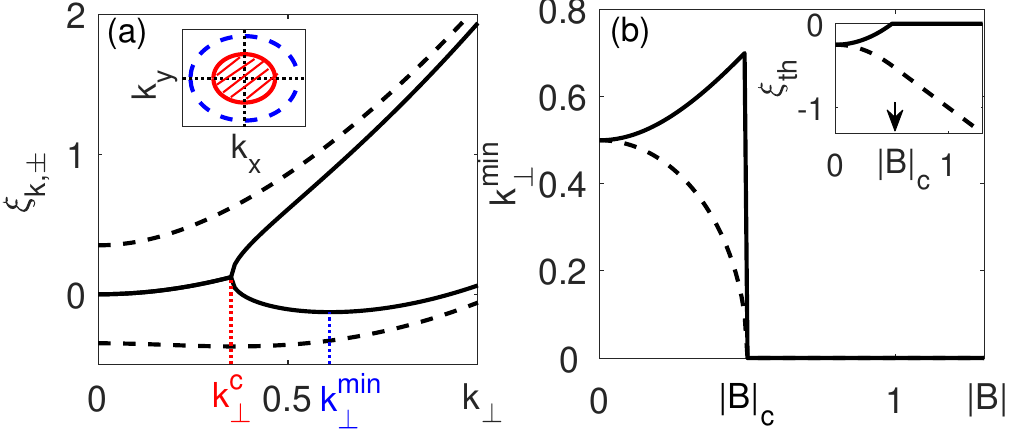}
\caption{(Color online). (a) Real parts of single-particle spectra under a Rashba SOC and an IMF (solid lines) or a RMF (dashed). Here we take $k_z=0$, and IMF/RMF with the same strength $|B|=0.35E_0$. 
Inset of (a): Exceptional ring (red solid, with radius $k_{\perp}^c$) and the location of energy minimum (blue dashed, with radius $k_{\perp}^{\rm min}$) in ($k_x,k_y$) plane in the case of   IMF. (b) Location of  energy minimum, $k_{\perp}^{\rm min}$, as a function of $|B|$ for IMF(solid) and RMF(dashed). Inset of (b) shows the corresponding threshold energy (in the real part) $\xi_{th}$. For IMF, $k_{\perp}^{\rm min}$ shows a discontinuity  and accordingly $\xi_{th}$ shows a kink at $|B|_c=0.5E_0$.  
In all plots, the units of momentum and energy are respectively $k_0$ and $E_0$. 
} \label{fig_spectrum}
\end{figure}

Eq.\ref{E_single} results in distinct single-particle spectra for IMF and RMF, as displayed in Fig.\ref{fig_spectrum}(a). For RMF, all $\xi_{\cp k,\pm}$ are real, and a gap is opened at ${\cp k}=0$; while for an IMF, $\xi_{\cp k,\pm}$ are complex (conjugate to each other) for $k_\perp<k^c_\perp\equiv |B|/\alpha$, and purely real for $k_\perp>k^c_\perp$. Right at $k_\perp=k^c_\perp$, both the two levels and two eigenstates coalesce, forming an exceptional ring in $(k_x,k_y)$ plane, see the red solid circle in the inset of Fig.\ref{fig_spectrum}(a). Note that since the size of this ring does not depend on $k_z$, in 3D ${\cp k}$-space it forms an exceptional surface as of a straight cylinder along $z$.
In Fig.\ref{fig_spectrum}(b), we plot the location of energy minimum, denoted by $k_{\perp}^{\rm min}$, as varying $|B|$. For RMF, $k_{\perp}^{\rm min}$ continuously decreases to zero as increasing $B$.  For IMF, $k_{\perp}^{\rm min}$ first increases with $|B|$ following $\sim (k_0/2)\sqrt{1+4|B|^2/E_0^2}$, as shown by the larger blue dashed circle in inset of Fig.\ref{fig_spectrum}(a), while at a critical $|B|_c=0.5E_0$ jumps to zero, signifying a first-order transition. Accordingly, at the transition point the energy threshold ($\xi_{th}$) moves from finite $k_{\perp}^{\rm min}$ to ${\cp k}=0$ and exhibits a kink, as shown in the inset of Fig.\ref{fig_spectrum}(b).

\begin{figure}[t]
\includegraphics[width=8.5cm]{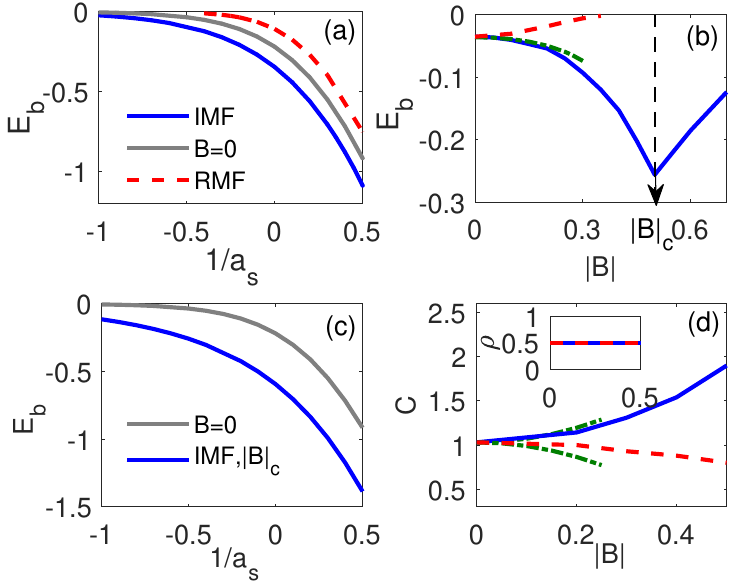}
\caption{(Color online). (a) Two-body binding energy $E_b$ with Rashba SOC as a function of $1/a_s$ for IMF(blue solid line) and RMF(red dashed) with $|B|/E_0=0.3$, in comparison to that with $B=0$(gray line). (b) $E_b$ as a function of $|B|$ for IMF(blue solid) and RMF(red dashed) given a fixed $1/(a_sk_0)=-0.5$. Green dashed-dot line shows  fit to Eq.\ref{E_b} for IMF ($B^2<0$). Under IMF, $E_b$ reaches minimum at  $|B|_c$. (c) Minimum $E_b$ under IMF (blue solid) as functions of $1/a_s$, in comparison to zero $B$ case (gray line). (d) Coupling constants $C$ between two threshold fermions as functions of $|B|$ for IMF(blue solid) and RMF(red dashed), and the inset shows threshold DoS $\rho$. Green dashed-dot lines show analytical fit(see text). 
} \label{fig_Eb}
\end{figure}

Now we come to the two-body problem, where two fermions interact under contact potential $U=g\delta({\cp r})P_{S=0}$, here ${\cp r}$ is the relative motion, $P_{S=0}$ is the projection operator of spin singlet state $|S=0\rangle=\frac{|\uparrow \downarrow\rangle-|\downarrow \uparrow \rangle}{\sqrt{2}}$, and the bare coupling $g$ can be related to the s-wave scattering length $a_s$ via $1/g=m/(4\pi a_s)-1/V\sum_{\cp k}1/2\epsilon_{\cp k}$. The two-body results are shown in Fig.\ref{fig_Eb}.

In Fig.\ref{fig_Eb}(a), we plot the two-body binding energy $E_b$ as a function of $1/a_s$ for IMF and RMF with the same amplitude $|B|$. In comparison to zero $B$ case, the application of an IMF will enhance $|E_b|$ at all couplings, while RMF always reduces $|E_b|$. The picture is more clearly shown in Fig.\ref{fig_Eb}(b), where  $E_b$ is plotted as a function of $|B|$ at given $1/a_s$. We can see that as increasing $|B|$ from zero, in RMF case the bound state quickly vanishes with $E_b\rightarrow 0$; while the IMF can support deeper bound state (with decreasing $E_b$) until $|B|$ reaches $|B|_c$, when the single-particle threshold $\xi_{th}$ displays a kink (see Fig.\ref{fig_spectrum}(b)). To highlight the dramatic effect of IMF in favoring bound states, in Fig.\ref{fig_Eb}(c) we plot the minimum $E_b$ (at $|B|=|B|_c$) as functions of $1/a_s$, in comparison to that without IMF. We can see that the IMF effect is visible in all interaction regime from weak to strong couplings. For instance, for a weak coupling $1/(a_sk_0)=-1$, at $|B|_c$ we have $|E_b|/E_0=0.115$, about twenty times larger than the value ($0.005$) at $B=0$. 

Physically, the enhanced bound state is associated with the IMF-increased coupling strength between low-energy states. To demonstrate this, we rewrite the two-body equation (see Methods) as
\begin{equation}
\frac{1}{g}=\int d\epsilon \rho(\epsilon)\frac{C(\epsilon)}{E-\epsilon}, \label{2body_2}
\end{equation}
where $\rho(\epsilon)$ and $C(\epsilon)$ respectively denote the density-of-state(DoS) and the coupling constant for two particles at scattering energy $\epsilon$. The change of these two values as varying $B$ directly determines the fate of bound state. It can be more transparently seen through in weak coupling limit, where the bound state formation is dominated by the low-energy scattering near $E\sim 2\xi_{th}$. In Fig.(\ref{fig_Eb}) (d) and its inset, we plot $C$ and $\rho$ at threshold $E=2\xi_{th}$ as varying $|B|$. We can see that while $\rho$ keeps static\cite{footnote_rho},  $C$ can increase (decrease) with $|B|$ in the case of IMF (RMF). Indeed, at small $B$, $C$ can be expanded as $C(B)=C(0)(1-4B^2/E_0^2)$, consistent with the numerical result shown in Fig.(\ref{fig_Eb}) (d). Then based on Eq.(\ref{2body_2}), we arrive at the following expansion of $E_b$ in weak coupling limit ($(k_0a_s)^{-1}\rightarrow -\infty$) and with small $B$ ($|B|\ll E_0$):
\begin{equation}
E_b(B)=E_b(0)\exp \left( -\frac{16B^2}{E_0^2} \frac{1}{k_0 |a_s|}\right), \label{E_b}
\end{equation}
here $E_b(0)$ is the binding energy at $B=0$.
Most remarkably, Eq.\ref{E_b} shows  that  by applying an IMF ($B^2<0$), $|E_b|$ can {\it exponentially} increase with a {\it huge} coefficient due to $1/(k_0 |a_s|)\gg 1$.  In contrast, applying a RMF ($B^2>0$) will exponentially reduce  $|E_b|$. We have confirmed that Eq.\ref{E_b} matches well with numerical results at small $|B|$, as shown by dashed-dot line in Fig.(\ref{fig_Eb}) (b).



Inspired by the two-body result, we now turn to the property of pairing superfluid for many fermions. Under the mean-field BCS theory, we introduce the paring order parameter $\Delta=(g/V)\sum_{\cp k} \ _L\langle c_{-{\cp k}\downarrow}c_{{\cp k}\uparrow}\rangle_R$ and $\tilde{\Delta}=(g/V)\sum_{\cp k} \ _L\langle c_{{\cp k}\uparrow}^{\dag}c_{-{\cp k}\downarrow}^{\dag}\rangle_R$, where $c_{{\cp k}\sigma}$ is the annilation operator of a free fermion of spin-$\sigma$ at ${\cp k}$, and $|\rangle_{R(L)}$ refers to the right (left) eigenvector for the BCS ground state. By some algebra, the thermodynamic potential $\Omega=H-\mu N$ can be diagonalized as
\begin{equation}
\Omega=\sum_{\cp k}' \left( \sum_{i=1}^4 E_{\cp k i}\alpha^{R \dagger}_{\cp k i}\alpha^{L}_{\cp k i}  +2(\epsilon_{\cp k}-\mu)\right)-\frac{\Delta\tilde{\Delta}}{g}. \label{Omega_2}
\end{equation}
Here $\alpha^{R \dagger}_{\cp k i}$ ($\alpha^{L}_{\cp k i}$) is the creation  (annilation) operator of the $i$-th right (left) quasi-particle with momentum ${\cp k}$, which satisfies the anti-commutation relation $\{\alpha^{R \dagger}_{\cp k i}, \alpha^{L}_{\cp k' j}\}=\delta_{{\cp k}{\cp k}'}\delta_{ij}$\cite{footnote_quasi}; the four quasi-particle energies follow
\begin{equation}
E_{\cp k}=\pm \sqrt{A_{\cp k}+B_{\cp k}\pm 2\sqrt{A_{\cp k}B_{\cp k} -(\alpha k_{\perp})^2 \Delta\tilde{\Delta}}}, \label{E_quasi}
\end{equation}
with $A_{\cp k}=(\epsilon_{\cp k}-\mu)^2+\Delta\tilde{\Delta},\ B_{\cp k}=\alpha^2 k_{\perp}^2+B^2$.
In contrast to the case of RMF where all $E_{\cp ki}$ are real, in the presence of an IMF they can be real or complex, depending on the values of ${\cp k}$ and other parameters $\Delta, \ \tilde{\Delta},\ \mu$ etc. Since only the product $\Delta\tilde{\Delta}$ matters  in the functional $\Omega$, but not individual $\Delta$ or $\tilde{\Delta}$, in the following we will choose a special case with $\Delta=\tilde{\Delta}$ and  minimize $\Omega$ (which is real) in terms of $\Delta$ to find the ground state.  

\begin{figure}[t]
\includegraphics[width=8.5cm]{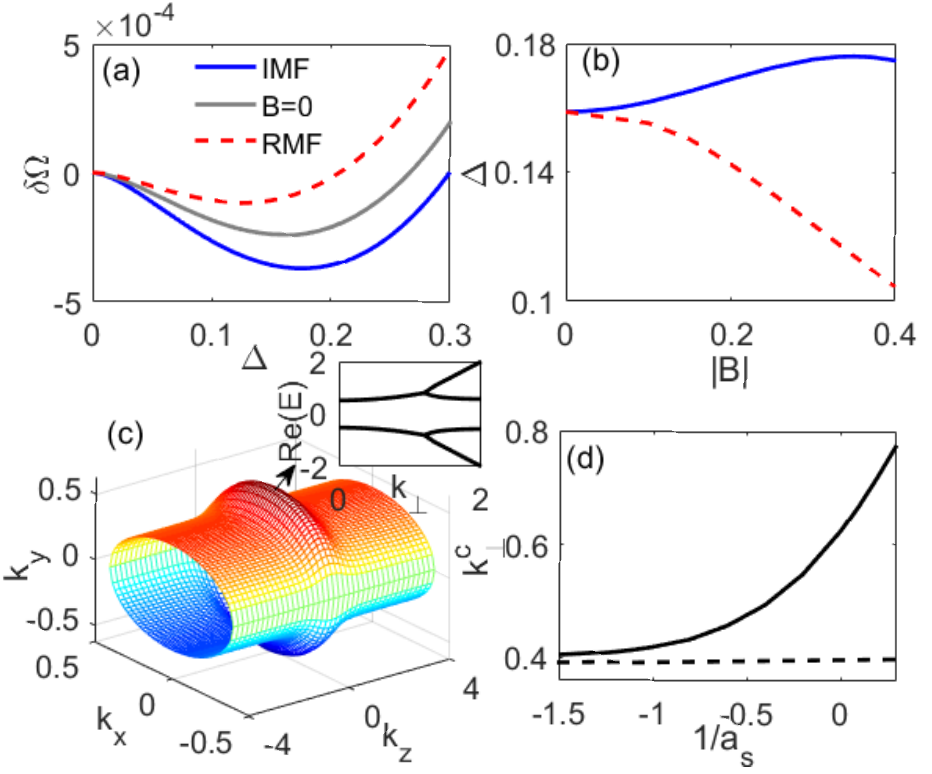}
\caption{(Color online). Fermion superfluid under Rashba SOC and IMF/RMF at a given $\mu=-0.05E_0$. (a)Thermodynamic potential $\Omega$ as a function of $\Delta$ for IMF(blue solid line) and RMF(red dashed) with $|B|=0.3E_0$, in comparison to that with $B=0$(gray line).  (b)Pairing amplitude $\Delta$ as varying $|B|$ for IMF(blue solid) and RMF(red dashed). In both (a,b), we take $1/(k_0a_s)=-0.5$. (c) ${\cp k}$-space exceptional surface for quasi-particles at resonance and with IMF strength $|B|=0.4E_0$. Inset shows the real parts of quasi-particle energies evolving with $k_{\perp}$ at fixed $k_z=0$, which split at $k_\perp^c$. (d) $k_\perp^c$ as a function of $1/a_s$ at given $|B|=0.4E_0$. Horizontal dashed lines shows $k_\perp^c$ for free particles.} \label{fig_sf}
\end{figure}

In Fig.\ref{fig_sf}(a), we show the typical landscapes of $\Omega(\Delta)$ for both IMF and RMF with a given strength $|B|$. It can be seen that compared to zero $B$ case, the IMF (RMF) can shift the minimum of $\Omega$ to larger (smaller) $\Delta$ and the according $\Omega_{\rm min}$ is further decreased (increased), indicating a stronger (weaker) fermion superfluid. In Fig.\ref{fig_sf}(b), we further plot the ground state $\Delta$, which indeed is an increasing function of $|B|$ for IMF, contrarily to the case of RMF. We have checked that these conclusions will not be qualitatively altered by the the change of $a_s$ and $\mu$.

In above, we have shown the dramatic effect of non-Hermitian potential (the IMF) to interacting fermions. In turn, the interaction effect can also alter the non-Hermitian property, in that the exceptional surface(ES) can be largely deformed from the free particle case. In Fig.\ref{fig_sf} (c), we show the ${\cp k}$-space ES of quasi-particles, which is determined by
\begin{equation}
A_{\cp k}B_{\cp k} =(\alpha k_{\perp})^2 \Delta\tilde{\Delta}.
\end{equation}
This equation predicts that two pairs of quasi-particles(\ref{E_quasi}) coalesce simultaneously at ES, with two separate energies $E_{\cp k}=\pm\sqrt{A_{\cp k}+B_{\cp k}}$,  as shown in the inset of Fig.\ref{fig_sf} (c). In comparison to the free particle case where ES is a straight cylinder along $z$ (see Fig.\ref{fig_spectrum}), here the ES can be deformed, as shown by Fig.\ref{fig_sf} (c), and the deformation is pronounced at low-energy space where the pairing takes a dominant role. As increasing the interaction strength, more and more momentum states will be strongly affected by pairing and ES will get even more distorted and extend to larger $k_{\perp}$, as manifested by the increasing $k_\perp^c$ for quasi-particles shown in Fig.\ref{fig_sf}(d).



To this end, we have demonstrated the enhanced pairing and superfluidity by an IMF for fermions with Rashba SOC. Next we show that such effect equally applies to other fermion systems, and in particular, we choose two types of  single-particle Hamiltonians as below:

(I) 
$h_0({\cp k})=\epsilon_{\cp k}\sigma_0+\alpha k_x\sigma_x +B\sigma_z$;

(II) 
$h_0({\cp k})=\epsilon_{\cp k}\sigma_0+\beta \sigma_x +B\sigma_z$

In comparison to the highly symmetric Rashba SOC as described by Eq.\ref{h0}, here we consider in case (I) a 1D SOC that has much less symmetry, and in case (II) a simple transverse field without any SOC.  Given their distinct structures, these three cases belong to the most typical situations for the non-trivial effect of IMF/RMF. In practice, case (I) with a real $B$ has been realized using the two-photon Raman process\cite{expt_1Dsoc}, and the transverse field in (II) and IMF in both (I,II) can be implemented respectively by the radio-frequency field and laser-assisted dissipation.

\begin{figure}[t]
\includegraphics[width=8.5cm]{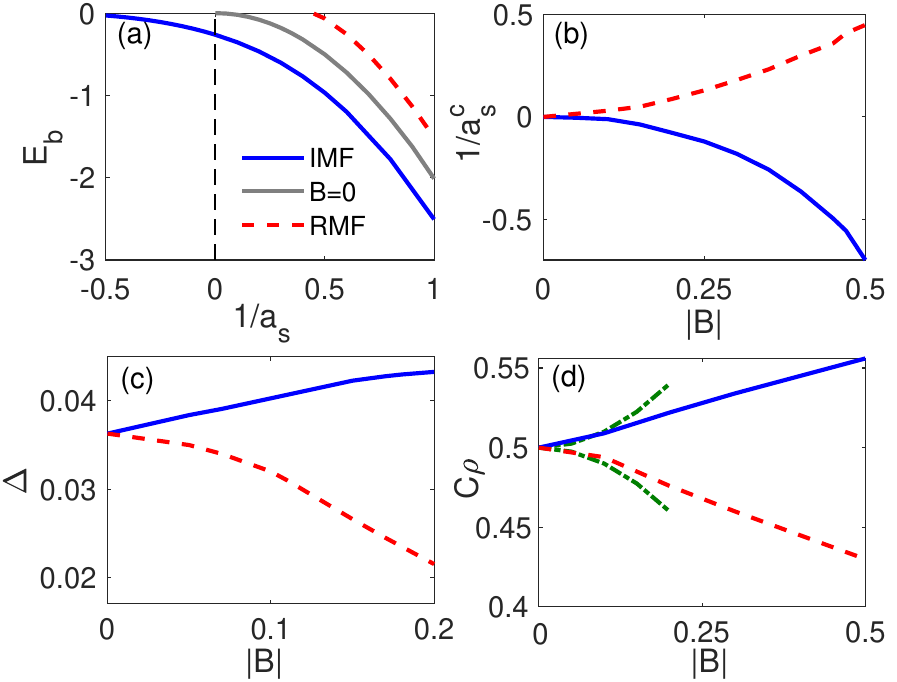}
\caption{(Color online). Pairing and superfluidity under 1D SOC with IMF/RMF. (a) Two-body binding energy $E_b$ as functions of $1/a_s$ for IMF(blue solid line) and RMF(red dashed) with $|B|=0.5E_0$, in comparison to that with $B=0$(gray line). (b)  Critical $1/a_s$ to support a two-body bound state as varying $|B|$ for IMF(blue solid) and RMF(red dashed).  (c) Many-body pairing amplitude $\Delta$ as a function of $|B|$ for IMF(blue solid) and RMF(red dashed). Here we take $1/(k_0a_s)=-0.5$ and $\mu=-0.05E_0$. (d) Product of coupling constant $C$ and DoS $\rho$ for two threshold fermions as varying $|B|$. Green dashed-dot lines show analytical fit (see text).} \label{fig_1dsoc}
\end{figure}

For case (I), we have carried out the two-body and many-body calculations for the pairing and superfluidity therein, which also show enhancement by  IMF, see Fig.\ref{fig_1dsoc}. 
In particular, with an IMF, the two-body bound state can form even in the weak coupling regime with $1/a_s<0$, on contrary to the case of RMF where it can only appear in the molecule side with $1/a_s>0$\cite{mol_1dsoc}, see Fig.\ref{fig_1dsoc} (a). In Fig.\ref{fig_1dsoc} (b), we plot the critical $1/a_s^c$ for the bound state formation as a function of $|B|$, and we see that the larger the IMF is, the weaker coupling (i.e., smaller $1/a_s^c$) is required to afford a bound state, while the RMF displays an opposite trend.  
Consistent with these two-body results, the many-body calculation shows the IMF-enhanced pairing amplitude $\Delta$, as contrast to the case of RMF, see Fig.\ref{fig_1dsoc} (c). The enhancement again can be understood from the analyses of low-energy coupling strength $C$ and DoS $\rho$. In this case, we have  $C(B)$ behaving the same as in Rashba SOC case while $\rho(B)$ varies as $\sim \rho(0)(1+2B^2/E_0^2)$, therefore $\rho(B)C(B)=\rho(0)C(0)(1-2B^2/E_0^2)$, which increases (decreases) with $|B|$ for an IMF (a RMF) as shown in Fig.\ref{fig_1dsoc} (d). Compare to the Rashba SOC case, here the magnitude of enhancement in 1D SOC is smaller due to the IMF-reduced DoS.

For case (II), the situation is much simpler, as the spin ($\sigma$) and orbit (${\cp k}$)  are fully decoupled in the single-particle level. Now  there is only one factor left, i.e., the spin gap, to influence the pairing and superfluidity properties. From the  spin spectrum  $\epsilon_{\pm}=\pm\sqrt{\beta^2+B^2}$, we define the spin gap as $G=\sqrt{\beta^2+B^2}$. 
Obviously, increasing $|B|$ in IMF and RMF cases will have different effects to $G$, i.e., in the former $G$ increases while in the latter $G$ decreases, until becoming zero at $|B|=\beta$  where locates the exceptional point. Such behavior can directly influence the many-body superfluidity. Indeed, the mean-field BCS theory gives quasi-particle spectra as:
\begin{equation}
E_{\cp k}=\pm \sqrt{(\epsilon_{\cp k}-\mu)^2+\Delta^2} \pm G,
\end{equation}
which shows that the spin gap $G$ directly plays the role of an effective magnetic field $h_{\rm eff}$ in  pairing problem. As it is known that changing $h_{\rm eff}$ can result in a sequence of quantum phase transitions between normal  and various pairing phases across resonances\cite{transitions},  such transitions can be equally induced  by changing the IMF strength $|B|$.  For trapped fermions, this effect directly leads to a tunable phase separation between normal and BCS pairing states, which can be measured directly as previously in spin-imbalanced Fermi gas \cite{imbalance_expt}.   

In conclusion, we have demonstrated the enhanced pairing and superfluidity by an IMF in a number of distinct fermion systems with and without SOC, by which we expect the same IMF effect can extend to a wide class of fermion systems. 
We have revealed the underlying mechanism for such enhancement as the IMF-increased low-energy coupling strength and IMF-reduced spin gap in forming singlet pairs. 
The remarkably opposite effects generated by an IMF and a RMF could be detected in cold atoms experiments. 
In particular, the binding energy of molecules can be measured by the rf spectroscopy, and the pairing superfluidity can be probed by the momentum-resolved rf spectroscopy\cite{JILA}.
Finally, while in this work we have only concentrated on the ground state property at zero temperature, our results immediately suggest an equally strong superfluid at finite temperature and with a high superfluid transition  temperature ($T_c$). This may also offer an alternative perspective towards the high-$T_c$ superconductor ever studied in literature.

{\bf Method.}

The two-body problem can be solved by using the Lippman-Schwinger equation $|\Psi \rangle=G_E U |\Psi\rangle$, where $\Psi$ is the two-body wave function. We then arrive at the following equation for binding energy $E_b=E-2\xi_{th}$:
\begin{equation}
 \frac{1}{g}=\langle S=0| G_E(0,0) |S=0\rangle, \nonumber
\end{equation}
where the Green function reads
\begin{equation}
G_E({\cp r},{\cp r'})=\frac{1}{2}\sum_{{\cp k};\mu\nu=\pm} \frac{\langle {\cp r}|{\cp k}^R_{\mu};-{\cp k}^R_{\nu}\rangle \langle-{\cp k}^L_{\nu};{\cp k}^L_{\mu}|{\cp r'}\rangle}{\langle {\cp k}^L_{\mu}|{\cp k}^R_{\mu}\rangle \langle -{\cp k}^L_{\nu}|-{\cp k}^R_{\nu}\rangle (E-\xi_{{\cp k}\mu}-\xi_{-{\cp k}\nu})} .  \nonumber
\end{equation}
Here $|{\cp k}^R_{\mu}\rangle$ and $|{\cp k}^L_{\mu}\rangle$ refer to the left and right eigenvectors defined through $h_0|{\cp k}^R_{\mu}\rangle=\xi_{{\cp k}\mu}|{\cp k}^R_{\mu}\rangle$ and $h_0^\dagger|{\cp k}^L_{\mu}\rangle=\xi^*_{{\cp k}\mu}|{\cp k}^L_{\mu}\rangle$. 
In principle, for the case of IMF the expansion in $G_E$ fails at the exceptional ring where there is only one eigenstate for each ${\cp k}$. Nevertheless, the integrand in $G_E$ behaves smoothly across the exceptional region, and thus its presence has no effect to the two-body solution. For the case of RMF, we have $h_0=h_0^\dagger$, $\xi_{{\cp k}\mu}=\xi^*_{{\cp k}\mu}$ and $|{\cp k}^R_{\mu}\rangle=|{\cp k}^L_{\mu}\rangle$.

For the many-body pairing, under the mean-field theory we can write the thermodynamic potential $\Omega$ as
\begin{equation}
\Omega=\sum_{\cp k}' \left( F^{\dagger}_{\cp k} \Omega({\cp k})F_{\cp k} +2(\epsilon_{\cp k}-\mu)\right)-\frac{\Delta\tilde{\Delta}}{g}, \nonumber
\end{equation}
with $F_{\cp k}=(c_{{\cp k}\uparrow}, c_{-{\cp k}\downarrow}^{\dagger}, c_{-{\cp k}\uparrow}^{\dagger},c_{{\cp k}\downarrow})^T$, and $\Omega({\cp k})$ is
\begin{equation}
\left(  \begin{array}{cccc}  \epsilon_{\cp k}-\mu+B    &  \Delta &0&\alpha k_\perp e^{-\i\phi_k} \\
 \tilde{\Delta} & -\epsilon_{\cp k}+\mu+B & \alpha k_\perp e^{-\i\phi_k} & 0\\
 0 & \alpha k_\perp e^{\i\phi_k} & -\epsilon_{\cp k}+\mu-B &-\tilde{\Delta}\\
  \alpha k_\perp e^{\i\phi_k} & 0& -\Delta & \epsilon_{\cp k}-\mu-B
 \end{array} \right).  \nonumber
\end{equation}
Note that  the summation over ${\cp k} $ in $\Omega$ is carried out only over half of ${\cp k}$-space. By diagonalizing the $4\times 4$ matrix at each ${\cp k}$, we can obtain the form of $\Omega$ as Eq.\ref{Omega_2} and the associated quasi-particle energy as Eq.\ref{E_quasi}.   

At zero temperature, we have $\Omega=\sum_{\cp k}' \left( \sum_{i=1}^4 E_{\cp k i}\Theta\big(-Re(E_{\cp k i})\big)  +2(\epsilon_{\cp k}-\mu)\right)-\Delta\tilde{\Delta}^2/g $, where $\Theta(x)=1$ if $x>0$ and $=0$ otherwise. The ground state of the system can be found by minimizing $\Omega$ as a function of the product $\Delta\tilde{\Delta}$, given $a_s, \ \mu, \ B$ all fixed. In the main text we have chosen a special case with $\Delta=\tilde{\Delta}$.

{\bf Acknowledgement.} We thank Wei Yi for helpful discussions. The work is supported by the National Key Research and Development Program of China (2018YFA0307600, 2016YFA0300603), and the National Natural Science Foundation of China (No.11622436, No.11421092, No.11534014).



\begin{thebibliography}{99}

\bibitem{FermiGas}B. DeMarco and D. S. Jin, Science 285, 1703 (1999); Giorgini, Pitaevskii, Stringari, Rev. Mod. Phys. {\bf 80}, 1215 (2008); M. W. Zwierlein, in Novel Superfluids (Oxford University Press, 2014) pp. 269-422.

%
\bibitem{2D}M. Feld, B. Fr{\" o}hlich, E. Vogt, M. Koschorreck, M. K{\" o}hl, Nature {\bf 480}, 75 (2011); A. T. Sommer, L. W. Cheuk, M. J. H. Ku, W. S. Bakr, and M. W. Zwierlein, Phys. Rev. Lett. {\bf 108}, 045302 (2012); P. A. Murthy,  M. Neidig, R. Klemt, L. Bayha, I. Boettcher, T. Enss, M. Holten, G. Z{\" u}rn, P. M. Preiss, S. Jochim,  Science {\bf 359}, 452 (2018).

\bibitem{narrow}E. L. Hazlett, Y. Zhang, R. W. Stites, and K. M. O'Hara, Phys. Rev. Lett. {\bf 108}, 045304 (2012); T.-L. Ho, X. Cui, and W. Li, Phys. Rev. Lett. {\bf 108}, 250401 (2012); R. Qi, H. Zhai, Phys. Rev. A {\bf 85}, 041603(R) (2012).

\bibitem{SOC}J. P. Vyasanakere, V. B. Shenoy, Phys. Rev. B {\bf 83}, 094515 (2011); J. P. Vyasanakere, S. Zhang, and V. B. Shenoy, Phys. Rev. B {\bf 84}, 014512 (2011); M. Gong, S. Tewari, and C. Zhang, Phys. Rev. Lett. {\bf 107}, 195303 (2011); H. Hu, L. Jiang, X. J. Liu, and H. Pu, Phys. Rev. Lett. {\bf 107}, 195304 (2011); Z. Q. Yu and H. Zhai, Phys. Rev. Lett. {\bf 107}, 195305 (2011); X. Cui, Phys. Rev. A {\bf 85}, 022705 (2012);  P. Zhang, L. Zhang, W. Zhang, Phys. Rev. A {\bf 86}, 042707 (2012); Y. Wu and Z. Yu, Phys. Rev. A {\bf 87}, 032703 (2013); S.-J. Wang and C. H. Greene, Phys. Rev. A {\bf 91}, 022706 (2015); Q. Guan, D. Blume, Phys. Rev. A {\bf 94}, 022706 (2016).

\bibitem{dissipation_expt}J. Li, A. K. Harter, J. Liu, L. de Melo, Y. N. Joglekar, and L. Luo, 
Nat. Comm. 855, 1 (2019); S. Lapp, J. Ang'ong'a, F. Alex An, B. Gadway, arxiv: 1811.06046.

\bibitem{Ueda-Yu} M. Nakagawa, N. Kawakami, M. Ueda, Phys. Rev. Lett. {\bf 121}, 203001 (2018); Z. Zhou, Z. Yu, arxiv: 1901.01174.

\bibitem{footnote_rho} $\rho$ is a constant at threshold,  due to the effective 2D DoS produced by the Rashba SOC.


\bibitem{footnote_quasi} Note that $\alpha^{R}_{\cp k i}$ and $\alpha^{L}_{\cp k i}$ are identical for RMF, but different for IMF.

\bibitem{expt_1Dsoc}
Y.-J. Lin, K. Jim\'{e}nez-Garc\'{i}a and I. B. Spielman, Nature {\bf471}, 83 (2011);
J.-Y. Zhang, S.-C. Ji, Z. Chen, L. Zhang, Z.-D. Du, B. Yan, G.-S. Pan, B. Zhao, Y.-J. Deng, H. Zhai, S. Chen and J.-W. Pan, Phys. Rev. Lett. {\bf109}, 115301 (2012);
P. Wang, Z.-Q. Yu, Z. Fu, J. Miao, L. Huang, S. Chai, H. Zhai and J. Zhang, Phys. Rev. Lett. {\bf109}, 095301 (2012);
L. W. Cheuk, A. T. Sommer, Z. Hadzibabic, T. Yefsah, W. S. Bakr and M. W. Zwierlein, Phys. Rev. Lett. {\bf109}, 095302 (2012);
C. Qu, C. Hamner, M. Gong, C. Zhang and P. Engels, Phys. Rev. A {\bf88}, 021604(R) (2013).


\bibitem{mol_1dsoc}R. A. Williams, M. C. Beeler, L. J. LeBlanc, K. Jimenez-Garcia, I. B. Spielman, Phys. Rev. Lett. {\bf 111}, 095301 (2013); L. Zhang, Y. Deng, and P. Zhang, Phys. Rev. A, {\bf 87}, 053626 (2012); D. M. Kurkcuoglu and C. A. R. Sá de Melo, Phys. Rev. A {\bf 93}, 023611 (2016). 

\bibitem{transitions}D. E. Sheehy and L. Radzihovsky, Phys. Rev. Lett. {\bf 96},
060401 (2006); C. H. Pao, S. T. Wu and S. K. Yip, Phys. Rev. B. {\bf 73},
132506 (2006); H. Hu and X. J. Liu, Phys. Rev. A. {\bf 73}, 051603(R) (2006); M. M. Parish, F. M. Marchetti, A. Lamacraft and B. D. Simons, Nature Physics {\bf 3}, 124 (2007).

\bibitem{imbalance_expt}M. W. Zwierlein, A. Schirotzek, C. H. Schunck, W. Ketterle,
Science {\bf 311}, 492 (2006); G. B. Partridge, W. Li, R. I. Kamar, Y. Liao, R. G. Hulet,
Science {\bf 311}, 503 (2006).

\bibitem{JILA} J. T. Stewart, J. P. Gaebler, D. S. Jin, Nature {\bf 454}, 744 (2007); J. P. Gaebler, J. T. Stewart, T. E. Drake, D. S. Jin, A. Perali, P. Pieri and G. C. Strinati, Nat. Phys. {\bf 6}, 569 (2010). 

\end{thebibliography}
\end{document}